\documentclass{webofc}
\usepackage[varg]{txfonts}   
\begin{document}
\title{Quarkonium production and polarization:\\where do we stand?}
%
%

\author{\firstname{Hee Sok}
\lastname{Chung}\inst{1}\fnsep\thanks{\email{neville@korea.ac.kr}}
}

\institute{Department of Physics, Korea University, Seoul 02841, Korea}

\abstract{%
We review the current status of heavy quarkonium
production phenomenology based on nonrelativistic effective field theories,
focusing on spin-triplet $S$-wave states such as $J/\psi$, $\psi(2S)$, and
$\Upsilon$. 
We present some representative examples for heavy quarkonium production 
mechanisms proposed in the literature, which vary significantly
depending on the choice of data employed in analyses. 
We then discuss the r\^{o}le of polarization in discriminating between the 
different possible scenarios for quarkonium production. 
Other observables that may be useful in pinpointing the production mechanism 
are also introduced, such as the $\eta_c$ production, associated production of 
$J/\psi$ plus a gauge boson, and $J/\psi$ production at the Electron-Ion
Collider. 
}
\maketitle
\section{Introduction}
\label{sec-1}

Heavy quarkonia are useful laboratories to study perturbative and
nonperturbative aspects of QCD~\cite{Brambilla:2004wf, Brambilla:2010cs,
Bodwin:2013nua, Brambilla:2014jmp}. An important class of observables involve 
inclusive production of heavy quarkonia, which are considered to be promising
contenders for tools to study QCD in colliders. This
requires a robust understanding of the quarkonium production mechanism based on
first principles, which remains a formidable challenge. 

Most theoretical studies of quarkonium production phenomenology rely on 
nonrelativistic effective field theories, which are based
on the fact that the mass $m$ of the heavy quark $Q$ and antiquark 
$\bar Q$ that constitute a quarkonium is much larger than 
$\Lambda_{\rm QCD}$. This allows an interpretation of heavy quarkonium states 
as nonrelativistic bound states; 
the scales that appear in a $Q \bar Q$ bound state are the momentum $mv$ and
the binding energy $mv^2$, where $m$ is the heavy quark mass and 
$v$ is the velocity of the $Q$ or $\bar Q$ inside
the quarkonium. Typical values of $v$ are $v^2 \approx 0.3$ for charmonia, 
and $v^2 \approx 0.1$ for bottomonia. The nonrelativistic QCD (NRQCD) 
effective field theory~\cite{Caswell:1985ui, Bodwin:1994jh}
provides a factorization formalism that separates the 
perturbative short-distance physics of scales of order $m$ and higher from
the nonperturbative long-distance physics which is encoded in the NRQCD matrix
elements. This formalism has been widely adopted in phenomenological studies 
of heavy quarkonium production. 

A difficulty in the NRQCD factorization approach is that, while the
short-distance part can be computed in perturbative QCD, it is generally not
known how to compute the long-distance quantities from first principles, and
they are usually determined phenomenologically. This approach has not lead to a
satisfactory description of the heavy quarkonium production
mechanism~\cite{Chung:2018lyq}.
Determinations of NRQCD matrix elements from different choices of data can
disagree with one another, and none of the determinations have been able to 
give a
comprehensive description of the important observables. Hence, it is fair to
say that a QCD-based understanding of the production mechanism of heavy
quarkonium still remains elusive. 

In this paper, we review recent efforts made towards understanding the heavy
quarkonium production mechanism based on nonrelativistic effective field
theories. 
In Sect.~\ref{sec-2}, we briefly introduce the nonrelativistic effective field
theory formalisms that are used in heavy quarkonium phenomenology. 
We review the various NRQCD matrix element determinations for 
spin-triplet $S$-wave
quarkonia, which include $J/\psi$, $\psi(2S)$, and $\Upsilon$ 
in Sect.~\ref{sec-3}. We discuss the r\^{o}le of quarkonium polarization in
understanding the production mechanism in Sect.~\ref{sec-4}, and 
introduce other observables related to quarkonium production in
Sect.~\ref{sec-5}. We conclude in Sect.~\ref{sec-6}.

\section{Nonrelativistic effective theories for quarkonium production}
\label{sec-2}
\subsection{Nonrelativistic QCD factorization}
\label{sec-2-1}

NRQCD provides a factorization formalism for inclusive production cross
sections of a heavy quarkonium $\cal Q$ in the form~\cite{Bodwin:1994jh}
\begin{equation}
\label{eq:NRQCDfac}
\sigma_{{\cal Q} +X} = \sum_n \sigma_{Q \bar Q(n) + X} 
\langle {\cal O}^{\cal Q} (n) \rangle, 
\end{equation}
where the sum is over color, spin, and orbital angular momentum states $n$ of
the $Q \bar Q$, $\sigma_{Q \bar Q(n) + X}$ are perturbatively calculable
short-distance cross sections for inclusive production of $Q \bar Q$ 
in the $n$ state, and 
$\langle {\cal O}^{\cal Q} (n) \rangle$ are 
NRQCD matrix elements that correspond to the nonperturbative probabilities for 
the $Q \bar Q (n)$ 
to evolve into a quarkonium $\cal Q$ plus anything. 
The matrix elements have known scalings in $v$, so that the sum over
$n$ in Eq.~(\ref{eq:NRQCDfac}) is organized in powers of $v$, and is in 
practice truncated at a desired order in $v$. 
The NRQCD matrix elements are universal quantities that depend on the
nonperturbative nature of the heavy quarkonium state. Hence,
Eq.~(\ref{eq:NRQCDfac}) provides
descriptions of inclusive quarkonium production rates with a few universal,
process-independent nonperturbative parameters. 

A novel feature of NRQCD factorization is that the quarkonium ${\cal Q}$ can be
produced from $Q \bar Q$ in color-octet states. In the case of $\chi_{cJ}$ and 
$\chi_{bJ}$ production, contributions at leading order in $v$ come from both 
color-singlet ($n={}^3P_J^{[1]}$, $J=0,1,2$) 
and color-octet ($n={}^3S_1^{[8]}$) channels. 
For production of spin-triplet $S$-wave quarkonia such as $J/\psi$, $\psi(2S)$,
and $\Upsilon$, the color-octet channel ($n={}^3S_1^{[8]}$,
$^1S_0^{[8]}$, $^3P_J^{[8]}$, $J=0,1,2$) 
contributions\footnote{Due to the heavy-quark spin symmetry, the $^3P_J^{[8]}$
channels for different $J$ involve the same NRQCD matrix element $\langle {\cal
O}^{\cal Q} (^3P_0^{[8]}) \rangle$. }
are suppressed by several powers of $v$ compared
to the contribution from the color-singlet channel ($n={}^3S_1^{[1]}$). 
However, when the transverse momentum $p_T$ of the quarkonium is much larger
than the heavy quarkonium mass, the short-distance cross sections for the
color-octet channels are strongly enhanced compared to the color-singlet
channel; moreover, the color-singlet channel contribution severely
underestimates the large-$p_T$ cross section measured at hadron colliders, 
so that the cross section is dominated by color-octet
contributions~\cite{Braaten:1994vv, Cho:1995vh, Cho:1995ce}. 
Hence, precise determination of the color-octet matrix elements is crucial for
understanding heavy quarkonium production based on first principles. 

It is worth mentioning that in general, NRQCD matrix elements are ultraviolet
divergent quantities that require renormalization. In particular, the 
color-octet matrix element $\langle {\cal O}^{\cal Q} (^3S_1^{[8]}) \rangle$ 
acquires dependence on the renormalization scheme and scale from one loop, 
in a way that 
contributions to the cross section from different channels mix
under renormalization. In $\chi_{cJ}$ and $\chi_{bJ}$ production, 
$^3P_J^{[1]}$ and $^3S_1^{[8]}$ channels mix under changes of the NRQCD scale, 
and in spin-triplet $S$-wave quarkonium production, 
$^3P_J^{[8]}$ and $^3S_1^{[8]}$ channels mix under
renormalization~\cite{Bodwin:1994jh, Bodwin:2012xc}. 
In calculations of the short-distance cross sections at one-loop level, the
matrix elements are usually renormalized in the $\overline{\rm MS}$
scheme at the scale of the heavy quark mass $m$. In this case, the 
short-distance cross sections for the $^3P_J^{[1]}$ and $^3P_J^{[8]}$ channels
are negative at values of $p_T$ much larger than the heavy quarkonium mass. 
Because of the mixing, only the sum of the contributions from all channels is
physically meaningful, while the contribution from a single channel can in
principle become negative. 

It is generally not known how to compute the NRQCD matrix elements
from first principles, except for the color-singlet matrix elements at leading
order in $v$, which can be related to decay matrix elements or quarkonium
wavefunctions at the origin~\cite{Bodwin:1994jh}. 
Because of this, the color-octet matrix elements
are usually determined phenomenologically by comparing Eq.~(\ref{eq:NRQCDfac})
with measured cross section data. 
As will be explained in a following section, the values of color-octet matrix
elements extracted from data depend strongly on the choice of measurements 
employed in the determination. 
This completely phenomenological approach to NRQCD matrix element
determination has not lead to a satisfactory description of the quarkonium
production mechanism, as none of the determinations have been able to give a
comprehensive description of important observables associated with inclusive
quarkonium production~\cite{Chung:2018lyq}. 
Some representative examples of matrix
element determinations will be shown in Sect.~\ref{sec-3}.

\subsection{Potential NRQCD}
\label{sec-2-2}

Recently, attempts have been made towards computing the NRQCD matrix elements
in the potential NRQCD (pNRQCD) effective field 
theory~\cite{Pineda:1997bj, Brambilla:1999xf, Brambilla:2004jw}. 
For strongly coupled quarkonia, pNRQCD provides
expressions for NRQCD matrix elements in terms of quarkonium wavefunctions at
the origin and universal gluonic correlators~\cite{Brambilla:2020ojz,
Brambilla:2021abf, Brambilla:2022rjd}. 
The gluonic correlators are defined by vacuum expectation values
of products of gluon field strengths and Wilson lines. 
A similar formalism has previously been used to compute NRQCD matrix elements
for quarkonium decays into light particles; in this case, 
the gluonic correlators have different definitions from the ones that appear 
in the production matrix elements~\cite{Brambilla:2002nu, Brambilla:2020xod}. 
In the case of color-singlet
matrix elements, the pNRQCD expressions reproduce at leading order in $v$ the
known results in terms of the wavefunctions at the origin. 
For the color-octet case, the gluonic correlators appear from leading order in
$v$ in the pNRQCD expressions for the matrix elements. 

Because the gluonic correlators do not involve heavy quark fields or projection
operators, they are universal quantities that do not depend on the specific 
heavy quarkonium state. In particular, the same gluonic correlators appear in
expressions for color-octet matrix elements for production of heavy quarkonia
with different radial excitation or heavy quark flavor. 
Based on this point, it has been argued that the gluonic correlators are more
basic quantities that are better suited for lattice QCD evaluations than the
original definitions for NRQCD matrix elements, although a lattice calculation 
of the correlators is yet to be
done~\cite{Brambilla:2020ojz, Brambilla:2021abf, Brambilla:2022rjd}. 

Even though first-principles determinations of NRQCD matrix elements through
lattice calculations of gluonic correlators have not been made possible yet, 
the pNRQCD expressions imply universal relations between color-octet matrix
elements for heavy quarkonium states with different radial excitation or heavy
quark flavor. 
Hence, the pNRQCD formalism allows
simultaneous inclusion of charmonium and bottomonium data in a single analysis
of color-octet matrix elements. 
This provides a strong constraint on color-octet matrix elements.
Results for NRQCD matrix element determinations in the pNRQCD formalism will be
shown in Sect.~\ref{sec-3}. 

\section{NRQCD matrix elements}
\label{sec-3}

We now list some representative examples of NRQCD matrix element determinations
based on calculations of the short-distance cross sections at next-to-leading
order in the strong coupling. 
The color-octet matrix elements for $J/\psi$ production from
refs.~\cite{Butenschoen:2011yh, Shao:2014yta, 
Zhang:2014ybe, Han:2014jya, 
Bodwin:2015iua, Feng:2018ukp, 
Brambilla:2022ayc} are shown in table~\ref{tab-jpsi}. 
In table~\ref{tab-psi2S} we show $\psi(2S)$ matrix elements from 
refs.~\cite{Shao:2014yta, Bodwin:2015iua, Feng:2018ukp, Brambilla:2022ayc, 
Butenschoen:2022orc}. For bottomonium, we show color-octet matrix elements for
production of $\Upsilon(1S)$, $\Upsilon(2S)$, and $\Upsilon(3S)$ states from
refs.~\cite{Gong:2013qka, Han:2014kxa, Brambilla:2022ayc} in
table~\ref{tab-upsilon}.

\begin{table}
\centering
\caption{Phenomenological determinations of $J/\psi$ color-octet matrix
elements in units of $10^{-2}$~GeV$^3$
from refs.~\cite{Butenschoen:2011yh, Shao:2014yta, 
Zhang:2014ybe, Han:2014jya, Bodwin:2015iua,
Feng:2018ukp, Brambilla:2022ayc}. In ref.~\cite{Shao:2014yta}, only two linear
combinations of the three matrix elements are determined, and the maximum
ranges of the matrix elements are obtained from maximizing and minimizing
$\langle {\cal O}^{J/\psi} (^1S_0^{[8]}) \rangle$ under the assumption made in
ref.~\cite{Shao:2014yta} that the matrix elements are positive definite.
Refs.~\cite{Bodwin:2015iua, Brambilla:2022ayc} provide covariance matrices from
which the correlations in the uncertainties can be obtained. 
The pNRQCD result from ref.~\cite{Brambilla:2022ayc} is from the analysis with 
$p_T$ larger than $5$ times the heavy quarkonium mass. 
}
\label{tab-jpsi}       
\begin{tabular}{l|ccc}
\hline
& 
$\langle {\cal O}^{J/\psi} (^3S_1^{[8]}) \rangle$ & 
$\langle {\cal O}^{J/\psi} (^1S_0^{[8]}) \rangle$ & 
$\langle {\cal O}^{J/\psi} (^3P_0^{[8]}) \rangle/m^2$  
 \\\hline
Ref.~\cite{Butenschoen:2011yh} 
& $0.168\pm 0.046$ & $3.04\pm0.35$ & $-0.404 \pm 0.072$ \\
Ref.~\cite{Shao:2014yta}, maximum $^1S_0^{[8]}$
& $0.05 \pm 0.02$ & $7.4 \pm 1.9$ & $0$
\\
Ref.~\cite{Shao:2014yta}, minimum $^1S_0^{[8]}$
& $1.1 \pm 0.3$ & $0$ & $1.9 \pm 0.5$  
\\
Ref.~\cite{Zhang:2014ybe} & $1.0 \pm 0.3$ & 0.44 -- 1.13 & $1.7 \pm 0.5$
\\
Ref.~\cite{Han:2014jya} & 0.9 -- 1.1 & 0 -- 1.46 & 1.5 -- 1.9
\\
Ref.~\cite{Bodwin:2015iua} & 
$-0.713 \pm 0.364$ & $11.0 \pm 1.4$ & $-0.312 \pm 0.151$
\\
Ref.~\cite{Feng:2018ukp} & $0.117 \pm 0.058$ & $5.66 \pm 0.47$ & 
$0.054 \pm 0.005$
\\
Ref.~\cite{Brambilla:2022ayc} & $1.40 \pm 0.42$ & $-0.63 \pm 3.22$ & $2.59 \pm
0.83$ \\
\hline
\end{tabular}
\end{table}

\begin{table}
\centering
\caption{
Phenomenological determinations of $\psi(2S)$ color-octet matrix
elements  in units of $10^{-2}$~GeV$^3$
from refs.~\cite{Shao:2014yta, Bodwin:2015iua, 
Brambilla:2022ayc, Butenschoen:2022orc}.
In ref.~\cite{Shao:2014yta}, only two linear combinations of the three matrix
elements are determined, and the maximum ranges of the matrix elements are
obtained from maximizing and minimizing $\langle {\cal O}^{\psi(2S)}
(^1S_0^{[8]}) \rangle$ under the assumption made in ref.~\cite{Shao:2014yta}
that the matrix elements are positive definite.  Refs.~\cite{Bodwin:2015iua,
Brambilla:2022ayc, Butenschoen:2022orc} provide covariance matrices from which
the correlations in the uncertainties can be obtained. 
The pNRQCD result from ref.~\cite{Brambilla:2022ayc} is from the analysis with 
$p_T$ larger than $5$ times the heavy quarkonium mass. 
Ref.~\cite{Butenschoen:2022orc} also provides results for fits including 
polarized cross sections, which are consistent with the results from including
only polarization-summed cross sections within uncertainties. 
}
\label{tab-psi2S}       
\begin{tabular}{l|ccc}
\hline
&
$\langle {\cal O}^{\psi(2S)} (^3S_1^{[8]}) \rangle$ &
$\langle {\cal O}^{\psi(2S)} (^1S_0^{[8]}) \rangle$ &
$\langle {\cal O}^{\psi(2S)} (^3P_0^{[8]}) \rangle/m^2$
\\ \hline
Ref.~\cite{Shao:2014yta}, maximum $^1S_0^{[8]}$
& $0.12 \pm 0.03$ & $2.0 \pm 0.6$ & $0$ \\
Ref.~\cite{Shao:2014yta}, minimum $^1S_0^{[8]}$
& $0.41 \pm 0.09$ & $0$ & $0.51 \pm 0.15$ \\
Ref.~\cite{Bodwin:2015iua} &
$-0.157 \pm 0.280$ & $3.14 \pm 0.79$ & $-0.114 \pm 0.121$ \\
Ref.~\cite{Brambilla:2022ayc} & 
$0.84 \pm 0.25$ & $-0.37 \pm 1.92$ & $1.55 \pm 0.49$ \\
Ref.~\cite{Butenschoen:2022orc}, $p_T > 1$~GeV
& $0.0537 \pm 0.0029$ & $1.00 \pm 0.03$ & $-0.218 \pm 0.005$ 
\\
Ref.~\cite{Butenschoen:2022orc}, $p_T > 7$~GeV 
& $0.225 \pm 0.025$ & $1.19 \pm 0.20$ & $0.272 \pm 0.053$
\\
\hline
\end{tabular}
\end{table}

\begin{table}
\centering
\caption{
Phenomenological determinations of $\Upsilon$ color-octet matrix
elements  in units of $10^{-2}$~GeV$^3$
from refs.~\cite{Gong:2013qka, Han:2014kxa, Brambilla:2022ayc}.
The analysis in ref.~\cite{Gong:2013qka} employed a smaller NRQCD scale 
of $1.5$~GeV whereas in refs.~\cite{Han:2014kxa, Brambilla:2022ayc} the scale 
was chosen to be the bottom quark mass $m_b$. Due to the running of the 
$^3S_1^{[8]}$ matrix element, at the scale $m_b$ 
the values of $\langle {\cal O}^{\Upsilon} (^3S_1^{[8]}) \rangle$ 
from ref.~\cite{Gong:2013qka} 
will be more negative for $\Upsilon(1S)$ and
$\Upsilon(2S)$, and more positive for $\Upsilon(3S)$ 
compared to what are listed in this table. 
It is also worth noting that the treatment of $P$-wave feeddowns in
ref.~\cite{Gong:2013qka} are inconsistent with measurements in
ref.~\cite{LHCb:2014ngh}, which became available after ref.~\cite{Gong:2013qka}
was published. 
In ref.~\cite{Han:2014kxa}, only two linear combinations of the three matrix
elements are determined, and the maximum ranges of the matrix elements are
obtained from maximizing and minimizing $\langle {\cal O}^{\Upsilon}
(^1S_0^{[8]}) \rangle$ under the assumption 
that the matrix elements are positive definite
similarly to what has been done in ref.~\cite{Shao:2014yta}. 
Ref.~\cite{Brambilla:2022ayc} provides covariance matrices from which the
correlations in the uncertainties can be obtained. 
The pNRQCD results from ref.~\cite{Brambilla:2022ayc} are from the analysis 
with $p_T$ larger than $5$ times the heavy quarkonium mass; the pNRQCD results
for $\Upsilon(1S)$ matrix elements are obtained by assuming the results in the 
strongly coupled pNRQCD formalism developed in refs.~\cite{Brambilla:2020ojz,
Brambilla:2021abf, Brambilla:2022rjd} also applies to the $1S$ state. }
\label{tab-upsilon}       
\begin{tabular}{l|ccc}
\hline
&
$\langle {\cal O}^{\Upsilon} (^3S_1^{[8]}) \rangle$ &
$\langle {\cal O}^{\Upsilon} (^1S_0^{[8]}) \rangle$ &
$\langle {\cal O}^{\Upsilon} (^3P_0^{[8]}) \rangle/m^2$
 \\\hline
Ref.~\cite{Gong:2013qka} $\Upsilon(1S)$ & 
$-0.41 \pm 0.24$ & $11.15 \pm 0.43$ & $-0.67 \pm 0.00$ \\
Ref.~\cite{Han:2014kxa} $\Upsilon(1S)$, maximum $^1S_0^{[8]}$
& $1.17 \pm 0.02$ & $13.7 \pm 1.11$ & $0$ \\
Ref.~\cite{Han:2014kxa} $\Upsilon(1S)$, minimum $^1S_0^{[8]}$
& $3.04 \pm 0.15$ & $0$ & $3.61 \pm 0.29$ \\
Ref.~\cite{Brambilla:2022ayc} $\Upsilon(1S)$ & 
$2.96 \pm 0.93$ & $-0.40 \pm 2.04$ & $2.12 \pm 0.68$ \\
\hline
Ref.~\cite{Gong:2013qka} $\Upsilon(2S)$ & 
$0.30 \pm 0.78$ & $3.55 \pm 2.12$ & $-0.56 \pm 0.48$ \\
Ref.~\cite{Han:2014kxa} $\Upsilon(2S)$, maximum $^1S_0^{[8]}$ &
$1.08 \pm 0.20$ & $6.07 \pm 1.08$ & $0$ \\
Ref.~\cite{Han:2014kxa} $\Upsilon(2S)$, minimum $^1S_0^{[8]}$ & 
$1.91 \pm 0.25$ & $0$ & $1.60 \pm 0.28$ \\
Ref.~\cite{Brambilla:2022ayc} $\Upsilon(2S)$ & 
$1.52 \pm 0.47$ & $-0.20 \pm 1.04$ & $1.08 \pm 0.35$ \\
\hline
Ref.~\cite{Gong:2013qka} $\Upsilon(3S)$ & 
$2.71 \pm 0.13$ & $-1.07 \pm 1.07$ & $0.39 \pm 0.23$ \\
Ref.~\cite{Han:2014kxa} $\Upsilon(3S)$, maximum $^1S_0^{[8]}$
& $0.83 \pm 0.02$ & $2.83 \pm 0.07$ & $0$ \\
Ref.~\cite{Han:2014kxa} $\Upsilon(3S)$, minimum $^1S_0^{[8]}$
& $1.22 \pm 0.02$ & $0$ & $0.74 \pm 0.02$ \\ 
Ref.~\cite{Brambilla:2022ayc} $\Upsilon(3S)$ & 
$1.17 \pm 0.37$ & $-0.16 \pm 0.81$ & $0.84 \pm 0.27$ \\
\hline
\end{tabular}
\end{table}

In all cases listed here, the color-singlet matrix elements employed in the
analyses are obtained from potential models or quarkonium decay rates, and are
consistent within uncertainties. 
The color-octet matrix elements are determined by
comparing Eq.~(\ref{eq:NRQCDfac}) with cross section data, 
taking into account the effect of feeddowns. 
We can see that the resulting values of the color-octet matrix elements differ
wildly, and even the signs of the matrix elements can be different,
although none of the color-octet matrix elements exceed the typical sizes 
expected from the nonrelativistic power counting: they are usually more than an
order of magnitude smaller than the color-singlet matrix element. 
With the exception of ref.~\cite{Butenschoen:2011yh}, 
the matrix element extractions are solely based on 
$p_T$-differential cross sections from hadron colliders 
with various choices of lower $p_T$ cuts. 
In these cases, an approximate degeneracy in the $p_T$ shapes of the
short-distance cross sections can prevent 
strongly constraining all three color-octet
matrix elements (see, e.g., ref.~\cite{Ma:2010jj}). 
In refs.~\cite{Shao:2014yta, Han:2014kxa}, 
only two linear combinations of the color-octet 
matrix elements were extracted, and the ranges of matrix elements were
determined by assuming positivity of the matrix elements; 
the results shown in the tables correspond to two extreme cases where 
$\langle {\cal O}^{\cal Q} (^1S_0^{[8]}) \rangle$ is maximized or minimized. 
Other
hadroproduction-based determinations from refs.~\cite{Bodwin:2015iua,
Brambilla:2022ayc, Butenschoen:2022orc} employed covariance-matrix
analyses to obtain linear combinations of matrix elements that are more suited
for phenomenological determinations; in many of these cases, one of the three
linear combinations is poorly determined compared to others, 
which corresponds to the undetermined linear
combination of matrix elements in refs.~\cite{Shao:2014yta, Han:2014kxa}. 
The hadroproduction-based approaches lead to predictions of the spin-triplet
$S$-wave quarkonium production mechanism that lie somewhere between two extreme
scenarios: in the $^1S_0^{[8]}$ dominance scenario, the cross
section is dominated by the $^1S_0^{[8]}$ channel contribution, while the 
sum of the $^3S_1^{[8]}$ and $^3P_0^{[8]}$ channel contributions are small; 
in the opposite scenario, the bulk of the cross section comes from the 
sum of the $^3S_1^{[8]}$ and $^3P_0^{[8]}$ channel contributions, while the 
$^1S_0^{[8]}$ contribution is small. In the hadroproduction-based approaches,
the $^3S_1^{[8]}$ and $^3P_0^{[8]}$ matrix elements have same signs, so that
the contributions from the two channels tend to cancel at large $p_T$, 
because there the short-distance cross sections for the two channels have 
opposite signs. 

The $J/\psi$ matrix elements in ref.~\cite{Butenschoen:2011yh} 
were obtained from a global fit of
cross section data including hadroproduction, photoproduction, and
$p_T$-integrated cross section at $B$ factories. 
This helps lift the approximate degeneracy in the $p_T$ shapes of the
short-distance cross sections, which allows all three color-octet matrix 
elements to be well determined. However, the values of the matrix elements
obtained in the global fit are
very different from hadroproduction-based approaches. In
refs.~\cite{Shao:2014yta, Zhang:2014ybe, Han:2014jya, Bodwin:2015iua,
Feng:2018ukp, Brambilla:2022ayc}, the
signs of the $^3S_1^{[8]}$ and $^3P_0^{[8]}$ matrix elements are same, leading
to cancellations between the two channels at large $p_T$. In contrast, 
in the global fit the $^3P_0^{[8]}$ matrix element is negative, while the
$^3S_1^{[8]}$ matrix element is positive, so that the contributions from the
two channels add at large $p_T$. Because of this, the $J/\psi$ hadroproduction
cross sections from the global fit tend to be in tension with measurements
at very large $p_T$. 

The $\psi(2S)$ matrix elements in ref.~\cite{Butenschoen:2022orc} 
were also obtained from a global fit of available cross section data; however,
unlike the $J/\psi$ case, availability of $\psi(2S)$ production data is mostly
limited to hadron collider experiments. 
The analysis with the cut $p_T > 1$~GeV shows a pattern of color-octet matrix 
elements that is similar to
the $J/\psi$ global fit, yielding a negative $^3P_0^{[8]}$ matrix element,
while the other two remain positive; this leads to predictions that are in
tension with measurements of the $p_T$ shape of the $\psi(2S)$ cross section,
as well as the polarization, as we will see in the next section. In contrast,
an alternative analysis with the cut $p_T > 7$~GeV presented in the same work
results in color-octet matrix elements that are 
similar to the $^3S_1^{[8]}$ plus $^3P_0^{[8]}$ dominance scenario 
from other hadroproduction-based approaches. 

The $J/\psi$ matrix element extractions in refs.~\cite{Zhang:2014ybe, 
Han:2014jya} are based on large-$p_T$ hadroproduction
data of $J/\psi$ and $\eta_c$ at the LHC. As will be explained in
Sect.~\ref{sec-5}, inclusion of the $\eta_c$ data gives additional constraints
to $J/\psi$ matrix elements through approximate heavy quark spin symmetry, 
and results in configurations where the $^1S_0^{[8]}$ channel contribution to
the $J/\psi$ production rate is small. 
That is, the analyses based on $J/\psi$ and $\eta_c$
hadroproduction data prefer the scenario where the sum of $^3S_1^{[8]}$ and 
$^3P_J^{[8]}$ channel contributions dominate the $J/\psi$ cross section. 

The pNRQCD-based analysis in ref.~\cite{Brambilla:2022ayc} 
employed hadroproduction data of $J/\psi$,
$\psi(2S)$, $\Upsilon(2S)$, and $\Upsilon(3S)$ at the LHC, by using the
universal relations between color-octet matrix elements for spin-triplet
$S$-wave quarkonia. This results in values of color-octet matrix elements 
that are better constrained than some conventional NRQCD approaches 
such as refs.~\cite{Shao:2014yta, Han:2014kxa}. 
This happens because the pNRQCD analysis includes the 
charmonium and bottomonium data simultaneously in the extraction of matrix 
elements, and acquires sensitivity to the running of the $^3S_1^{[8]}$ matrix
element at scales ranging from the charm to the bottom quark masses. 
Because the one-loop anomalous dimension of the $^3S_1^{[8]}$ matrix element is
proportional to the $^3P_0^{[8]}$ matrix element, this constrains the
$^3P_0^{[8]}$ matrix element to a positive definite value.  As a result, the
pNRQCD analysis yields a configuration of color-octet matrix elements where the
bulk of the cross sections come from the sum of $^3S_1^{[8]}$ and $^3P_J^{[8]}$
channel contributions for all spin-triplet $S$-wave quarkonia. 

As have been shown in this section, the phenomenological determinations of 
color-octet matrix elements result in values that vary wildly depending on the
choice of data. 
Notably, the large-$p_T$ analyses based on $J/\psi$ and
$\eta_c$ hadroproduction data~\cite{Zhang:2014ybe, Han:2014jya}, 
as well as the pNRQCD analysis based on
charmonium and bottomonium hadroproduction data~\cite{Brambilla:2022ayc},
favor the $^3S_1^{[8]}$ plus $^3P_J^{[8]}$ dominance scenario, 
in contrast with global fits including low-$p_T$ data and 
other hadroproduction-based approaches favoring 
$^1S_0^{[8]}$ dominance.

\section{\boldmath Polarization of $J/\psi$, $\psi(2S)$, and $\Upsilon$ in 
hadron colliders}
\label{sec-4}

The polarization of spin-triplet $S$-wave heavy quarkonia has long been
considered an important test of the color-octet matrix elements. 
Early analyses based on tree-level calculations of the short-distance cross
sections predicted that the $J/\psi$ will be strongly transverse at large
$p_T$~\cite{Leibovich:1996pa, Beneke:1996yw, Braaten:1999qk}. 
This has not been supported by experiment: measurements at the LHC 
show little or no evidence of any strong polarization of 
spin-triplet $S$-wave quarkonia (see for example refs.~\cite{CMS:2012bpf, 
CMS:2013gbz}). 

The tree-level prediction of transversely polarized $J/\psi$ was based on the
observation that only the $^3S_1^{[8]}$ channel can contribute appreciably at
large $p_T$. This no longer holds at one loop: all three color-octet channels 
can contribute at large $p_T$ through gluon fragmentation~\cite{Gong:2008sn,
Gong:2008hk, Gong:2008ft, Butenschoen:2010rq, Ma:2010yw, Ma:2010jj,
Gong:2012ug}. The polarization can still discriminate between different 
color-octet channels, because the polarization of the quarkonium is affected by
the spin and orbital angular momentum of the $Q \bar Q$ produced in gluon 
fragmentation. For both $^3S_1^{[8]}$ and
$^3P_J^{[8]}$ channels, the transverse polarization of the fragmenting gluon is
mostly transferred to the $Q \bar Q$, because the fragmentation can occur by
emitting soft gluons: as a result, the $Q \bar Q$ produced in $^3S_1^{[8]}$ and
$^3P_J^{[8]}$ channels is mostly transverse, while the longitudinal production
rate is small; note that, due to the subtraction of the infrared divergence,
the large-$p_T$ transverse production rate of $Q \bar Q(^3P_J^{[8]})$ is
negative, while the longitudinal production rates are positive.  On the other
hand, the $^1S_0^{[8]}$ channel is isotropic, so it cannot produce polarized
final states. 

Unpolarized spin-triplet $S$-wave quarkonia can be produced in two ways: if the
production rate is dominated by the $^1S_0^{[8]}$ channel, then the quarkonium
cannot be strongly polarized, because $Q \bar Q(^1S_0^{[8]})$ is isotropic. 
In the $^3S_1^{[8]}$ plus $^3P_J^{[8]}$ dominance scenario, the color-octet
matrix elements for the two channels have same signs, 
so the transverse production
rate largely cancels between the two channels, while the longitudinal cross
sections add; this way, unpolarized final states can be obtained even when the 
$^1S_0^{[8]}$ channel contribution is small. 
In contrast, the global fit analyses with small lower $p_T$ cuts that give
negative values for the $^3P_0^{[8]}$ matrix elements yield transversely
polarized quarkonia at large $p_T$, because the transverse production rates
from the $^3S_1^{[8]}$ and $^3P_J^{[8]}$ channels add. 
As a result, the polarization measurements at the LHC are in tensions with the
predictions based on the global fit analyses with small lower $p_T$
cuts~\cite{Butenschoen:2011yh, Butenschoen:2022orc}, 
while the polarization results based on large-$p_T$ hadroproduction measurements
agree with experiments~\cite{Shao:2014yta,
Zhang:2014ybe, Han:2014jya, Han:2014kxa, Bodwin:2015iua,
Feng:2018ukp, Brambilla:2022ayc}. 

A shortcoming of the use of polarization for discriminating color-octet matrix
elements is that it can hardly distinguish between the $^1S_0^{[8]}$ dominance
and the $^3S_1^{[8]}$ plus $^3P_J^{[8]}$ dominance scenarios, because both
cases lead to unpolarized quarkonia. In the case of the pNRQCD analysis, which
favors the $^3S_1^{[8]}$ plus $^3P_J^{[8]}$ dominance scenario, $\Upsilon$ is
predicted to be more transverse than $J/\psi$ or $\psi(2S)$ due to the running
of the $^3S_1^{[8]}$ matrix element coming from the large and positive 
$^3P_J^{[8]}$ matrix element; this running would not have a prominent 
effect to polarization in the $^1S_0^{[8]}$ dominance scenario. 
Even though this prediction agrees with measurements of $\Upsilon$ polarization
at the LHC~\cite{CMS:2012bpf}, which show slightly more
transverse polarization than $J/\psi$, this effect is numerically
small and diluted by feeddown effects, especially for $1S$ and $2S$
bottomonia. This makes it desirable to have more observables that may help
distinguish between the two competing scenarios.

\section{Comparison with other observables}
\label{sec-5}

There have been quite a few observables related to inclusive quarkonium 
production measured in collider experiments, but many have not been able to
strongly scrutinize the heavy quarkonium production mechanism. 
For example, the Belle measurement for the total inclusive $J/\psi$ production 
rate~\cite{Belle:2009bxr} involves an unknown branching fraction into four or 
more charged tracks;
furthermore, it is unclear whether the form of NRQCD factorization given in
Eq.~(\ref{eq:NRQCDfac}) would hold for total inclusive production
rates\footnote{One reason would be that in such case, the $Q$ and $\bar Q$ 
do not necessarily need to be produced within a distance of $1/m$ 
in order to produce a quarkonium, which
would not allow for the usual form of NRQCD factorization to hold.}. 
In the case of photoproduction, measurements at the DESY HERA~\cite{H1:2002voc, H1:2010udv} were made with
kinematical cuts on the elasticity, which can make it difficult for NRQCD to
make reliable predictions~\cite{Beneke:1998re}. 
Studies of $J/\psi$ momentum distribution in jet~\cite{Bain:2016clc, 
Bain:2017wvk} showed that the measured
distribution from LHCb~\cite{LHCb:2017llq} 
is incompatible with the global fit results for $J/\psi$
matrix elements, while the matrix elements in the $^1S_0^{[8]}$ dominance 
scenario lead to results that are in fair agreement with measurements. 
Although a calculation based on the $^3S_1^{[8]}$ plus $^3P_J^{[8]}$ dominance
scenario has not been done in ref.~\cite{Bain:2017wvk}, 
we can expect that this will yield results that are
qualitatively similar to the $^1S_0^{[8]}$ dominance scenario, based on the
general behavior of the shapes of the $^3S_1^{[8]}$ and $^3P_J^{[8]}$ channel
contributions to the distribution.

There are still several observables proposed in the literature 
that can help distinguish the different 
scenarios for quarkonium production mechanism. 
The $\eta_c$ production rate measured by LHCb~\cite{LHCb:2014oii, LHCb:2019zaj} 
has been considered a good
observable, as it gives additional constraints for $J/\psi$ matrix elements 
based on heavy quark spin symmetry. Heavy quark spin symmetry implies that the
$J/\psi$ and $\eta_c$ matrix elements that differ by one unit of the $Q \bar Q$
spin are same at leading order in $v$, up to calculable spin multiplicity 
factors. That is, the $^1S_0^{[8]}$ matrix element for $J/\psi$ determines the 
$^3S_1^{[8]}$ matrix element for $\eta_c$. 
In the case of $\eta_c$, the cross section is dominated by $^1S_0^{[1]}$ and 
$^3S_1^{[8]}$ channels, so that the measured cross section gives a strong
constraint on the $^1S_0^{[8]}$ matrix element for
$J/\psi$~\cite{Zhang:2014ybe, Han:2014jya, Butenschoen:2014dra}. 
The measurements imply that 
the $^1S_0^{[8]}$ contribution to the $J/\psi$ cross section must be small, 
because the color-singlet contribution makes up for the bulk of the 
measured $\eta_c$ production rate. 
As a result, the analyses based on 
$J/\psi$ and $\eta_c$ production data favor the $^3S_1^{[8]}$ plus
$^3P_J^{[8]}$ dominance scenario, as have been presented in the previous
section. 
Similarly, the pNRQCD analysis leads to $\eta_c$ production rates that are 
compatible with measurements, albeit with large uncertainties due to the
limited precision for $\langle {\cal O}^{J/\psi} (^1S_0^{[8]}) \rangle$. 
A shortcoming of the NRQCD description of the $\eta_c$ production rate
currently adopted in the literature is that,
unlike the $J/\psi$ case, the contribution from the color-singlet channel at 
leading order in $v$ is significant; recall that, in the case of $S$-wave
quarkonia, the color-octet matrix elements are suppressed by several
powers of $v$ compared to the color-singlet one. 
This means that in the $\eta_c$ case, it may be necessary that the relativistic
corrections to the color-singlet channel must be included up to relative
order $v^4$, because they can be the same order as the color-octet 
contributions. 
The tension between measurement and NRQCD calculations of exclusive
production rates of $\eta_c$ at $B$ factories, which only involve color-singlet
contributions, may imply that the relativistic corrections to the color-singlet
channel can be significant~\cite{Chung:2008km, Sang:2009jc,Li:2009ki,
Fan:2012dy, Xu:2014zra, Belle:2018jqa, Chung:2019ota}. 
This effect has so far not been taken into account in existing analyses of
inclusive $\eta_c$ production. 

Another observable that may help discriminate the quarkonium production
mechanism is the associated production of a heavy quarkonium plus a gauge
boson. Calculations of short-distance cross sections at one-loop level have been
done for the production of $J/\psi$ plus a photon~\cite{Li:2014ava}, 
and the weak gauge bosons $W$ and $Z$~\cite{Butenschoen:2022wld}.  
Measurements have been made available by ATLAS 
for production of $J/\psi+W$~\cite{ATLAS:2014yjd, ATLAS:2019jzd} and
$J/\psi+Z$~\cite{ATLAS:2014ofp}. 
The data are mostly available for $J/\psi$ transverse momentum
larger than the $J/\psi$ mass.  
A recent analysis from ref.~\cite{Butenschoen:2022wld} shows that only the 
$^3S_1^{[8]}$ plus $^3P_J^{[8]}$ dominance scenario\footnote{The results from
the pNRQCD analysis in ref.~\cite{Brambilla:2022rjd} was shown in
ref.~\cite{Butenschoen:2022wld} as a representative case.}
results in predictions 
for the $J/\psi+W$ and $J/\psi+Z$ production rates that are compatible with 
measurements; the $^1S_0^{[8]}$ dominance scenario can even lead to negative
direct cross sections, and the global fit that mainly
comes from low-$p_T$ data gives cross sections that underestimate
data. 

Finally, predictions for the $J/\psi$ production cross sections from 
electron-proton collisions at the Electron-Ion Collider have recently been 
made available~\cite{Qiu:2020xum}. 
It has been shown that the large-$p_T$ hadroproduction-based
analyses lead to predictions that are distinct from what is obtained from the 
global fit~\cite{Qiu:2020xum, Brambilla:2022ayc}. 
While the $^3S_1^{[8]}$ plus $^3P_J^{[8]}$ dominance scenario
yields slightly larger $p_T$-differential production rates than the 
$^1S_0^{[8]}$ dominance scenario, precise measurements of the cross sections at
large $p_T$ will be needed to distinguish between the two scenarios.

\section{Summary and outlook}
\label{sec-6}

In this paper we have presented a concise review of the current status of
phenomenology of inclusive heavy quarkonium production and polarization based
on nonrelativistic effective field theories. 
Theoretical calculations of heavy quarkonium production rates in the
nonrelativistic QCD (NRQCD) factorization formalism require perturbative 
calculations of the short-distance cross sections as well as nonperturbative 
determinations of NRQCD matrix elements. 
While color-singlet matrix elements have been computed in potential models
and lattice QCD or determined from decay rates, color-octet matrix elements
have not been computed from first principles. 
In the case of the production of spin-triplet $S$-wave quarkonia,
color-octet matrix elements for the $^3S_1^{[8]}$, $^1S_0^{[8]}$, and the 
$^3P_J^{[8]}$ channels have significant contributions to the cross section. 

While perturbative QCD calculations of short-distance cross sections have been
carried out at one-loop accuracy for many important processes including
hadroproduction and polarization at the LHC, 
results for phenomenological determinations of 
NRQCD matrix elements depend strongly on the choice of data. 
For $J/\psi$ and $\psi(2S)$, large-$p_T$ hadroproduction-based 
determinations lead to scenarios where the cross section is dominated by 
either the $^1S_0^{[8]}$ channel or the remnant of the cancellation between 
$^3S_1^{[8]}$ and $^3P_J^{[8]}$ channels that mix under renormalization. 
On the other hand, global fits that include data with $p_T$ similar or smaller
than the heavy quarkonium mass result in values of color-octet matrix elements 
that make $^3S_1^{[8]}$ and $^3P_J^{[8]}$ channel contributions add at 
large $p_T$. 

The matrix element determinations from global fits including low-$p_T$ 
data~\cite{Butenschoen:2011yh, Butenschoen:2022orc}
and large-$p_T$
hadroproduction-based analyses~\cite{Gong:2013qka, Shao:2014yta,Han:2014kxa, 
Zhang:2014ybe, Han:2014jya, Bodwin:2015iua,
Feng:2018ukp, Brambilla:2022ayc} lead to contrasting predictions for
polarization. Because both the $^3S_1^{[8]}$ and $^3P_J^{[8]}$ channels are
strongly transversely polarized, global fits including low $p_T$ data 
predict transversely polarized charmonia at large $p_T$, while
hadroproduction-based approaches predict almost no polarization. 
Polarization measurements at the LHC 
disfavor the low-$p_T$ global fit predictions, showing no strong
evidence of polarization. While LHC polarization measurements seem to agree
with predictions from the hadroproduction-based analyses, 
polarization cannot strongly discriminate between the 
$^1S_0^{[8]}$ dominance and $^3S_1^{[8]}$ plus $^3P_J^{[8]}$ dominance
scenarios, because they both lead to similar near-zero polarization
predictions. 

This unfavorable situation could be improved by efforts from both theory 
and experiment. On the theory side, the potential NRQCD (pNRQCD) effective
field theory has been employed to further factorize the NRQCD matrix elements
into quarkonium wavefunctions at the origin and universal gluonic
correlators~\cite{Brambilla:2020ojz,
Brambilla:2021abf, Brambilla:2022rjd, Brambilla:2022ayc}. 
While first-principles determinations of NRQCD matrix elements through lattice
calculations of the gluonic correlators are yet to be done, the universality
of the gluonic correlators give rise to relations between color-octet matrix
elements for different heavy quarkonium states, which provide additional
constraints in phenomenological extractions of matrix elements. 
Analyses based on pNRQCD calculations of the color-octet matrix elements 
and large-$p_T$ hadroproduction data favor
the $^3S_1^{[8]}$ plus $^3P_J^{[8]}$ dominance scenario for all spin-triplet
$S$-wave quarkonium states including $J/\psi$, $\psi(2S)$, and $\Upsilon$. 
On the experimental side, measurements of additional observables such as 
$\eta_c$ production~\cite{LHCb:2014oii, LHCb:2019zaj} and the 
associated production of $J/\psi+W$~\cite{ATLAS:2014yjd, ATLAS:2019jzd} 
and $J/\psi+Z$~\cite{ATLAS:2014ofp} 
at the LHC have also been shown to prefer the $^3S_1^{[8]}$ plus $^3P_J^{[8]}$
dominance scenario~\cite{Zhang:2014ybe, Han:2014jya, Butenschoen:2022wld}. 

While it looks promising that analyses based on large-$p_T$ production seem to
be converging to the $^3S_1^{[8]}$ plus $^3P_J^{[8]}$ dominance scenario, 
it is well known that these approaches lead to bad
descriptions of low-$p_T$ observables, including total inclusive production
rates in lepton colliders and photoproduction cross sections at
HERA~\cite{Butenschoen:2012qr}. Even in hadroproduction, analyses based on
large $p_T$ production have trouble describing low-$p_T$ data, as has been
demonstrated in ref.~\cite{Butenschoen:2022orc}. 
The fact that the heavy quarkonium production mechanism that correctly 
describes both high and low $p_T$ regions still remains out of reach suggests 
that there is much more to be understood in QCD and factorization formalisms.

\begin{acknowledgement}

This work is supported by Korea University and by 
the National Research Foundation of Korea (NRF) Grant funded by the Korea
government (MSIT) under Contract No. NRF-2020R1A2C3009918.

\end{acknowledgement}

\bibliography{hsc_confxv}

\end{document}